\documentclass[journal=nalefd,manuscript=article,english]{achemso}

\usepackage{chemformula} 
\usepackage[T1]{fontenc} 
\usepackage{amsmath,amssymb,amsfonts,mathrsfs}
\usepackage{amsthm}
\usepackage{CJK}
\usepackage{bm}
\usepackage{babel}
\usepackage{graphicx}
\usepackage{txfonts}
\let\mathbb=\varmathbb
\DeclareSymbolFont{letters}{OML}{ztmcm}{m}{it}

\author{Yuanjun Jin}
\affiliation{Department of Physics $\&$ Institute for Quantum Science and Engineering, Southern University of Science and Technology,
Shenzhen 518055, P. R. China.}
\author{Rui Wang}
\email{rcwang@cqu.edu.cn}
\affiliation{Department of Physics $\&$ Institute for Quantum Science and Engineering, Southern University of Science and Technology,
Shenzhen 518055, P. R. China.}
\alsoaffiliation{Institute for Structure and Function $\&$
Department of physics, Chongqing University, Chongqing 400044, P. R. China.}
\author{Hu Xu}
\email{xuh@sustc.edu.cn}
\phone{+86(0)755 88018210}
\fax{+86(0)755 88018210}
\affiliation{Department of Physics $\&$ Institute for Quantum Science and Engineering, Southern University of Science and Technology,
Shenzhen 518055, P. R. China.}

\title[An \textsf{achemso} demo]
  {Recipe for Dirac Phonon States with Quantized Valley Berry Phase in Two Dimensional Hexagonal Lattices}

\begin{document}


\newpage
\begin{abstract}
The topological quantum states in two-dimensional (2D) materials are fascinating subjects of research, which usually highlight electron-related systems. In this work, we present a recipe that leads to Dirac phonon states with quantized valley Berry phase in 2D hexagonal lattices by first-principles calculations. We show that candidates possessing the three-fold rotational symmetry at the corners of the hexagonal Brillouin zone host valley Dirac phonons, which are guaranteed to remain intact with respect to perturbations.
We identify that such special topological features populated by Dirac phonons can be realized in various 2D materials.
In particular, the monolayer CrI$_3$, an attractive 2D magnetic semiconductor with exotic applications in spintronics, is an ideal platform to investigate nontrivial phonons in experiments.
We further confirm that the phonon Berry phase is quantized to $\pm \pi$ at two inequivalent valleys. The phonon edge states terminated at the projection of phonon Dirac cones are clearly visible.
This work demonstrates that 2D hexagonal lattices with attractive valley Dirac phonons will extend the knowledge of valley physics, providing wide applications of topological phonons.
\end{abstract}

~~\\
~~\\
\textbf{KEYWORDS:} Dirac phonons, valley Berry phase, phonon edge states, 2D hexagonal lattices, first-principles calculations

\newpage
The discovery of topological quantum states is one of the most promising advancements in condensed matter physics. Up to now, topological concepts have mainly highlighted electron-related systems, such as topological insulators\cite{Kane-RevModPhys.82.3045, ZSC-RevModPhys.83.1057, 3dqahewang} or topological semimetals.\cite{RevModPhys.90.015001, PhysRevB.84.235126, PhysRevB.97.195157, Wang216} These findings immensely strengthen the understanding of band topology and low-energy excitations of electrons in crystalline solids.  Topological orders in electronic materials provide potentially intriguing applications of dissipationless electronic devices or topological quantum computation.\cite{RevModPhys.88.021004} Beyond the various nontrivial fermionic electrons, the investigations of topological states are also intensively generalized to bosonic systems, such as photons in photonic crystals\cite{PhysRevLett.100.013904, PhysRevA.78.033834, PhysRevLett.100.013905, Naturephotonic2013}, classical elastic waves in macroscopic artificial phononic crystals,\cite{Natcom2015, Natphys2016, Science2015, Natphys2018, PhysRevB.97.035442} and phonons in crystalline materials.\cite{PhysRevB.96.064106, PhysRevLett.120.016401, PhysRevLett.121.035302, PhysRevB.97.054305, mgbphonon} The subjects of nontrivial bosons greatly enrich the physics of symmetry-protected topological orders. Of particular importance is topological phonons induced by atomic vibrations at THz frequency, which play critical roles in thermal transports, electron-phonon coupling, or multi-phonon process. However, there are only a few materials related to topological phonons to date, such as double Weyl phonons in transition-metal monosilicides,\cite{PhysRevLett.120.016401,PhysRevLett.121.035302} two-component and three-component phonons in WC-type compounds,\cite{PhysRevB.97.054305} and nodal-line phonons in MgB$_{2}$.\cite{mgbphonon} All these reported candidates possess three-dimensional (3D) linear dispersions in their phonon band structures. In comparison with 3D cases, the symmetries are lower in 2D crystals. Therefore, 2D materials with less symmetry constraints can more intuitively exhibit topological features of phonons. Unfortunately, the 2D phonon topology in realistic materials has been rarely reported in the literatures.

To investigate the phonon topological features in 2D, we recall the electronic band topology in graphene.\cite{RevModPhys.81.109} The discovery of graphene has given rise to tremendous interests in 2D materials.\cite{PhysRevLett.105.136805, PhysRevLett.108.155501, ACSNano, NatRM} More importantly, its electronic band structure possessing two degenerate and inequivalent valleys in momentum space exhibits topological semimetallic features in the absence of spin-orbital coupling (SOC).\cite{PhysRevLett.99.236809} The massless Dirac fermions are protected by space-time inversion ($\mathcal{IT}$) with spin-rotation symmetry.\cite{PhysRevLett.116.156402}  However, SOC is a fundamental effect of electrons in solids, and thus Dirac cones at the corners of the hexagonal Brilllouin zone (BZ) in graphene must be destroyed, converting into a quantum spin Hall (QSH) insulator with a nontrivial gap (i.e., a 2D topological insulator).\cite{PhysRevLett.95.226801}  A phonon is the quantized elementary vibrational mode of interacting atoms, which is typically spinless\cite{Levine1962}. Hence, the degenerate bands of Dirac phonons are not hybridized due to the lack of SOC, and the time-reversal ($\mathcal{T}$) symmetry of phononic systems is generally conserved. Recently, theoretical investigations indicated that phonons in hexagonal lattices could exhibit pseudospin-related topological effects and valley phonon Hall effects.\cite{PhysRevLett.119.255901, PhysRevLett.115.115502, NCR} Therefore, it is highly desirable to explore or design topological phonon states in 2D realistic candidates with the hexagonal symmetry.

\begin{figure}
	\centering
	\includegraphics[scale=0.35]{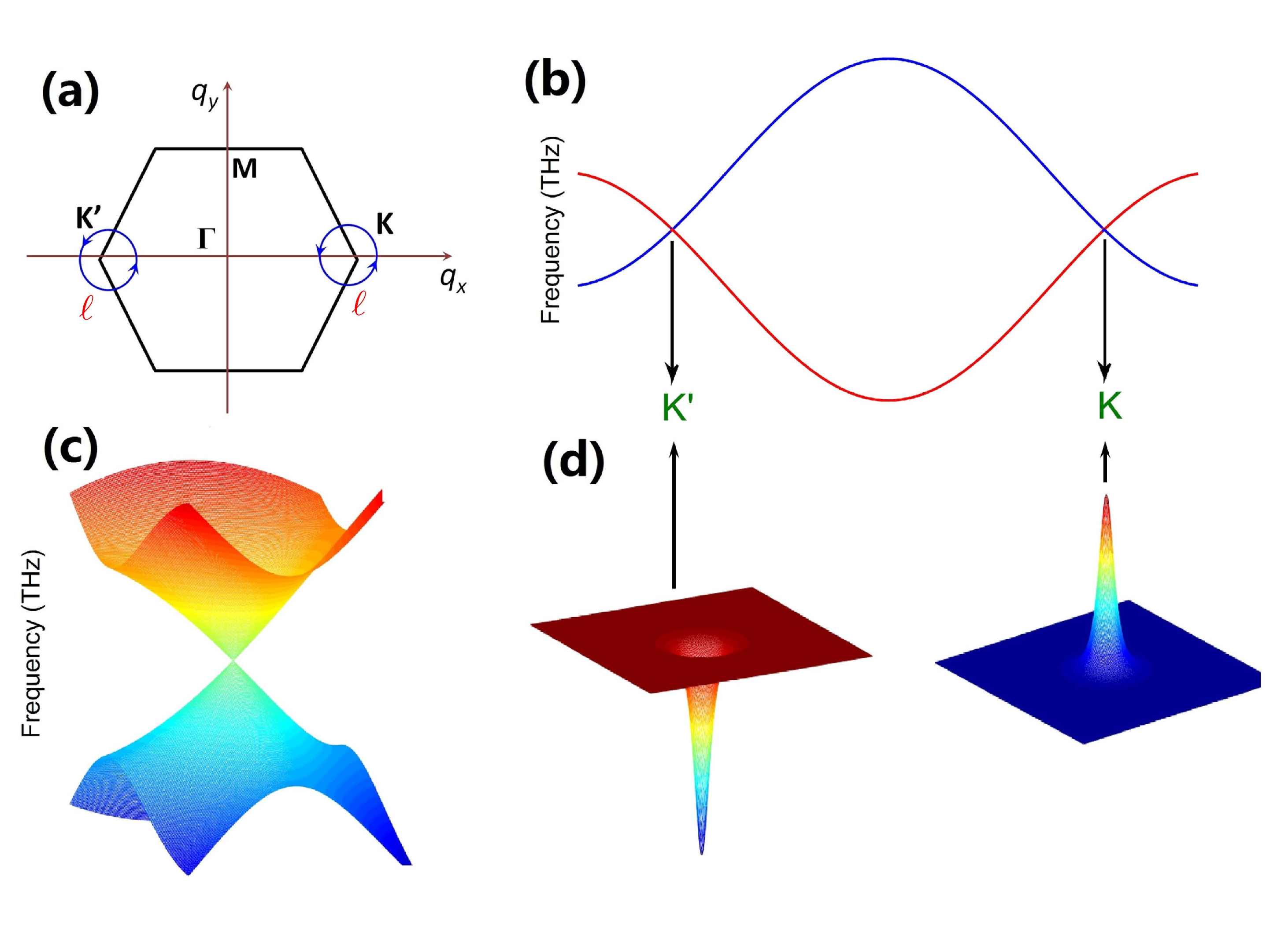}
	\caption{ Schematic of topological Dirac phonon states in 2D hexagonal lattices. (a) The hexagonal BZ with the high-symmetry points labeled. The two distinct valleys are denoted by $K$ and $K'$, respectively. A closed path $\ell$ with locally gapped spectra encircling a valley center is marked to calculate the phonon Berry Phase. (b) The Dirac phonons are formed by linear band crossings between two arbitrary adjacent phonon branches at $K$ and $K'$ points. (c) A 3D representation of the isotropic phonon Dirac dispersion. (d) The distributions of phonon Berry curvature near two inequivalent valleys $K$ and $K'$, respectively, indicating that a phonon Dirac point is a singularity in momentum space.
\label{figure-berry}}
\end{figure}

We first elucidate the topological features of phonons in hexagonal lattices.  Considering $\mathcal{IT}$ symmetry and three-fold rotational $C_3$ symmetry, many 2D materials crystallized in hexagonal lattices can possess linear crossings in phonon spectra at the corners of BZ, i.e., the $K$ and $K'$ points as shown in \ref{figure-berry}(a). The $K$ and $K'$ points also respectively represent two distinct valleys. Phonons are typically bosons, and therefore the topological features in the whole range of phonon frequencies can be detected. Here, we consider the Dirac phonon states formed by two arbitrary adjacent phonon branches that linearly cross at the $K$ and $K'$ points [see \ref{figure-berry}(b)]. Since the high-symmetry points $K$ and $K'$ possess the $C_3$ symmetry, the phonon Dirac cones are isotropic as shown in \ref{figure-berry}(c). The quasiparticles near the Dirac points can be described by an effective Hamiltonian as
\begin{equation}
H=v_D(q_x\tau_{z}\sigma_x+q_y \sigma{y}),
\end{equation}
where $H$ is referenced to the frequency of a Dirac point, $v_D$ is the corresponding phonon group velocity, $\sigma_{i}$ are Pauli matrices, $\tau_{z}=\pm 1$ respectively label two distinct valleys $K$ and $K'$, and $\mathbf{q}$ represents the wave vector relative to the valley centers.  It is well known that the Berry phase is useful to reveal topological features.  Similar to  electron systems, we can also define the phonon Berry connection
\begin{equation}\label{Berryc}
\mathbf{A}(\mathbf{q})=-i\sum_{\lambda}{\langle \varphi_{\lambda}(\mathbf{q})|\nabla_{\mathbf{q}}|\varphi_{\lambda}(\mathbf{q})\rangle},
\end{equation}
where $\varphi_{\lambda}(\mathbf{q})$ is the  Bloch wavefunction of the $\lambda$th phonon branch. Combining the lattice dynamical matrix $D(\mathbf{q})$ and the atomic eigen-displacement $\mathbf{u}_{\mathbf{q}}$, the phonon Bloch wavefunction is expressed as\cite{PhysRevB.96.064106,NCR}
\begin{equation}\label{wavefunction}
\varphi_{\lambda}(\mathbf{q}) = \left(
                                  \begin{array}{c}
                                    D_{\mathbf{q}}^{1/2} \mathbf{u}_{\mathbf{q}} \\
                                    \mathbf{\dot{u}}_{\mathbf{q}} \\
                                  \end{array}
                                \right),
\end{equation}
where $\mathbf{\dot{u}}_{\mathbf{q}}$ is time derivative of $\mathbf{u}_{\mathbf{q}}$. \ref{Berryc} and \ref{wavefunction} indicate that we can calculate the Berry connection $\mathbf{A}(\mathbf{q})$ using atomic force constants, which can be obtained from a finite displacement method\cite{PhysRevLett.78.4063} or density-functional perturbation theory (DFPT).\cite{RevModPhys.73.515} The corresponding phonon Berry curvature is
\begin{equation}
\mathbf{\Omega}(\mathbf{q})=\nabla \times \mathbf{A}(\mathbf{q})={\Omega}_z(\mathbf{q})\mathbf{e}_{z}.
\end{equation}
Since the phonon Dirac point is a singularity in momentum space, $\mathbf{\Omega}(\mathbf{q})$ must be divergent at this point. Besides, the presence of $\mathcal{IT}$ symmetry guarantees $\mathbf{\Omega}(\mathbf{q})$ with  opposite values at the inequivalent valleys  $K$ and $K'$  [see \ref{figure-berry}(d)].  For a closed path  $\ell$ with locally gapped spectra encircling a valley center [see \ref{figure-berry}(a)], we can define the valley Berry phase as $\gamma_{\ell} = \oint_{\ell} \mathbf{A}(\mathbf{q}) \cdot d\mathbf{q}$, which must be quantized in units of $\pi$.

\begin{figure}
	\centering
	\includegraphics[scale=0.35]{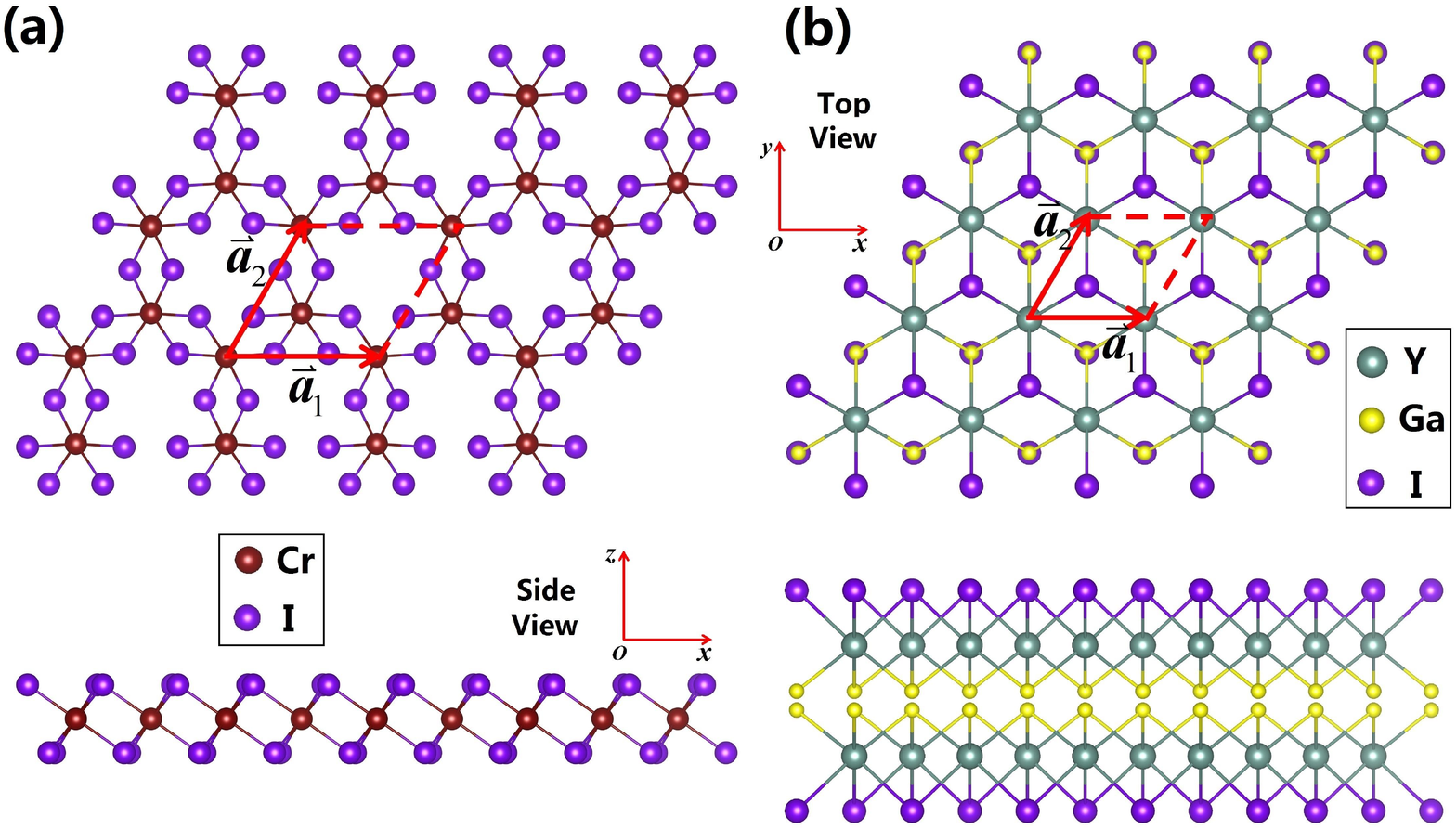}
	\caption{The atomic Crystal structures in hexagonal lattices. Top (upper panels) and side (lower panels) views of the monolayer (a) CrI$_3$ and (b) YGaI, respectively. Red rhombuses indicate the primitive cell with two unit primitive cell vectors $\mathbf{a}_1$ and $\mathbf{a}_2$.
\label{figure-struct}}
\end{figure}

The connection to Berry-related properties allows us to reveal the topological features of phonons. In this work, we present a recipe that leads to Dirac phonon states with quantized valley Berry phase in 2D hexagonal lattices with inversion ($\mathcal{I}$) symmetry. We explain our recipe using numerical calculations of the quantized Berry phase and topological edge states.  Based on this recipe, we show that 2D candidates with different chemical formulas and compositions can exhibit such special properties. In the main text, we use the monolayer CrI$_3$ and YGaI as examples to illustrate the topologically nontrivial features of phonons. Detailed results of other materials in our recipe are included in Supporting Information (SI). In particular, the recent magneto-optical Kerr effect microscopy demonstrated that the monolayer CrI$_3$ is an attractive 2D magnetic semiconductor and exhibits Ising ferromagnetism below a Curie temperature ($T_c$) of 61 K with out-of-plane spin orientation.\cite{nature2017CrI, Chemmeter, JiangNano}

To verify the existence of phonon topology in our recipe, we combined high-throughput and lattice dynamics calculations to obtain phonon spectra with fixing the Bravais lattices in 2D hexagonal symmetry. We carried out first-principles calculations using the Vienna \textit{ab initio} Simulation Package (VASP)\cite{Kresse2} based on density functional theory.\cite{Kohn} The exchange-correlation potential employed Perdew-Burke-Ernzerhof type generalized gradient approximation.\cite{Perdew1,Perdew2} The core-valence interactions were treated by the projector augmented wave  method\cite{Kresse4,Ceperley1980} with the cutoff energy of 600 eV. An accurate optimization of structural parameters was calculated by minimizing the forces on each atom smaller than 0.0001 eV/{\AA}. The full BZ was sampled by $21\times21\times1$ Monkhorst-Pack grid.\cite{Monkhorst} Real-space force constants were generated from DFPT with $4\times4\times1$ supercells. The phonon dispersions were calculated by diagonalization of force constants as implemented in the PHONOPY code.\cite{TOGO20151} To reveal the phonon topological features, we constructed a Wannier tight-binding (TB) Hamiltonian of phonons from the second rank tensor of force constants.\cite{TOGO20151} Then, the phonon Berry phase can be determined from integrals of phonon Berry curvature using the Wilson-loop method.\cite{Yu2011,Soluyanov2011} The iterative Green's function method was employed to obtain phonon edge states of semi-infinite nanoribbons.\cite{Sancho1984}

As shown in \ref{figure-struct}(a) and \ref{figure-struct}(b), the 2D candidates CrI$_3$ and YGaI both crystallize in hexagonal lattices with primitive unit cell vectors
$\mathbf{a}_1=a\mathbf{\hat{x}}$ and $\mathbf{a}_2=\frac{1}{2}a\mathbf{\hat{x}}+\frac{\sqrt{3}}{2}a\mathbf{\hat{y}}$, where $a=|\mathbf{a}_1|=|\mathbf{a}_2|$ is the lattice constant.  Then two distinct valleys $K$ and $K'$ are centered at $\mathbf{q}=(\pm 4\pi/3a,0)$, respectively. The optimized lattice constants $a$ for CrI$_3$ and YGaI are 6.96 {\AA} and 4.21 {\AA}, respectively. For CrI$_3$, we can see that the Cr atomic layer forms a honeycomb lattice, which is sandwiched between two layers of I atoms arranged on triangular-planar lattices. As a result, each Cr atom is surrounded by six I atoms and each I atom bonds to two Cr atoms.  For YGaI, Y, Ga, and I atoms are respectively composed as two individual layers of honeycomb lattices, exhibiting a stacking structure. Two Y atomic layers are stacked in AA configuration, and each Y atom connects two I atoms and two Ga atoms. Two Ga atomic layers stacked in AB configuration are in the middle, and two I atomic layers are respectively terminated at the upper and lower sides.

\begin{figure}
	\includegraphics[scale=0.58]{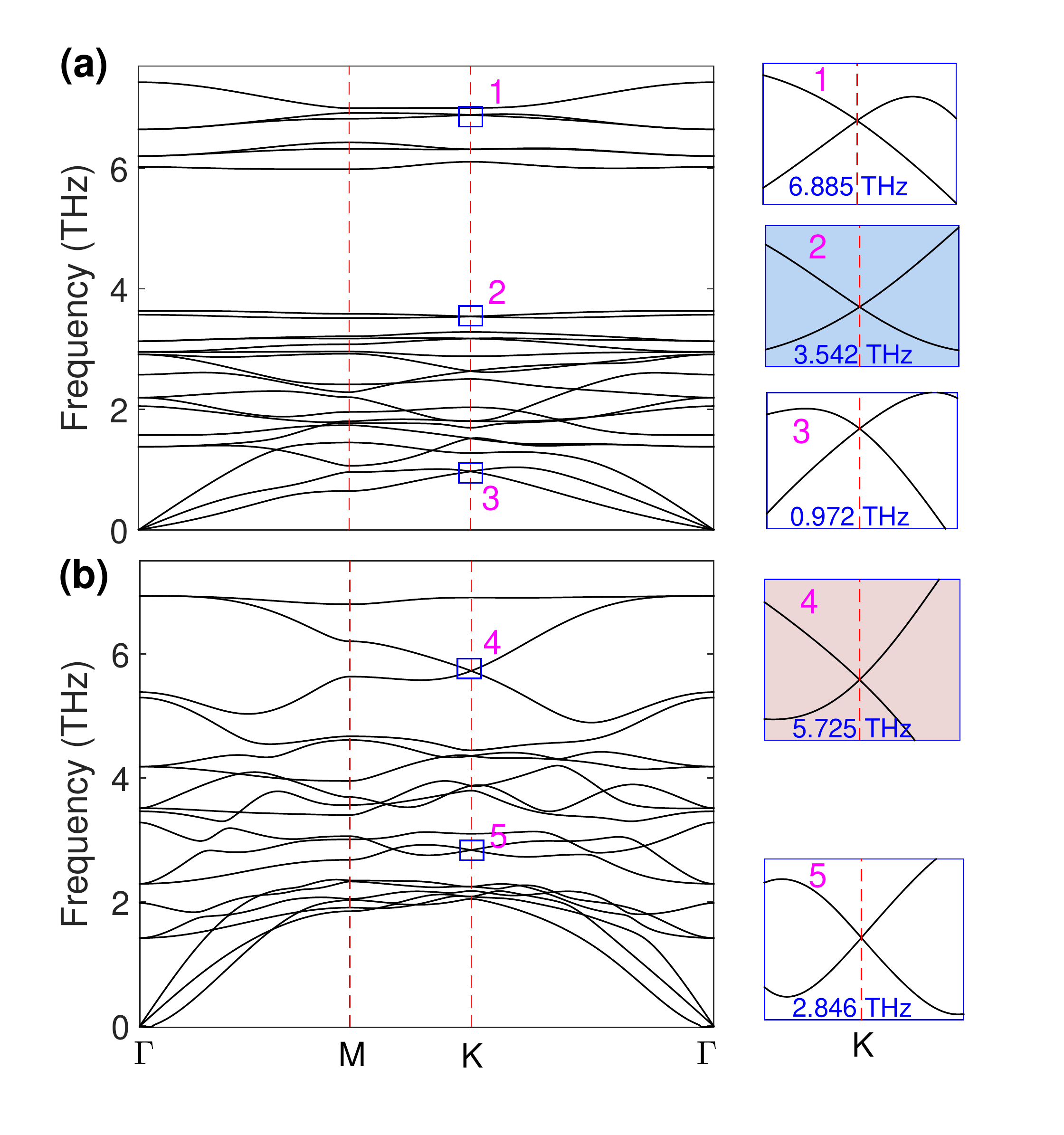}
	\caption{The phonon spectra of (a) CrI$_3$ and (b) YGaI along high-symmetry directions. The enlarged view illustrated in the right panels show that the linear band crossings at $K$ (or $K'$) points are visible, and the Dirac frequencies are also labeled. The dispersions near the ideal Dirac cones of CrI$_3$ ($\omega = 3.542$ THz) and YGaI ($\omega = 5.725$ THz) are highlighted in the panels 2 and 4, respectively.
\label{figure-phonon}}
\end{figure}

The phonon spectra of CrI$_3$ are shown in  \ref{figure-phonon}(a). As expected, we can see that there are three visible phonon band crossings at the $K$ (or $K'$) point. Of particular importance is the band crossing at the frequency $\omega = 3.542$ THz, for which the two nontrivial crossing branches are well separated from trivial branches. The ideal Dirac phonon states can be easily detected and are important to applications of topological quantum transport of phonons. Similar to CrI$_3$, the linear Dirac phonon dispersions at the $K$ (or $K'$) point in YGaI are also visible [see  \ref{figure-phonon}(b)], and there is also an ideal Dirac cone at the frequency $\omega = 5.725$ THz.

The presence of band degeneracies at the $K$ (or $K'$) points of hexagonal BZ is due to the coexistence of $\mathcal{{IT}}$ symmetry for phonons, even though some candidates in our recipe may be magnetic. For instance, the monolayer CrI$_3$ possesses the ferromagnetic order below $T_c$. The magnetic structures can break the $\mathcal{{T}}$ symmetry of a crystal, especially for electrons. However, phonons are typically neutral bosons\cite{Levine1962} and do not couple to magnetic fields (or magnetization) to the first order. Besides, there is no structural phase transition between ferromagnetic and nonmagnetic phases of CrI$_3$.\cite{nature2017CrI}  In our theoretical frame, the magnetic order originated from electrons only modifies the amplitudes of the force constants between interacting atoms. As the force constants are always real, the high-order coupling between phonons and magnetization can be ignored.  To confirm this, we make a comparison between the ferromagnetic and nonmagnetic phases of the phonon spectra of CrI$_3$ (see Figure S4 in SI). There are only tiny frequency shifts, implying that the topological Dirac phonon features in the the ferromagnetic and nonmagnetic phases are essentially the same. Thus, the $\mathcal{{T}}$ symmetry breaking for phonons in CrI$_3$ is so weak and can be negligible.

\begin{figure}
	\centering
	\includegraphics[scale=0.35]{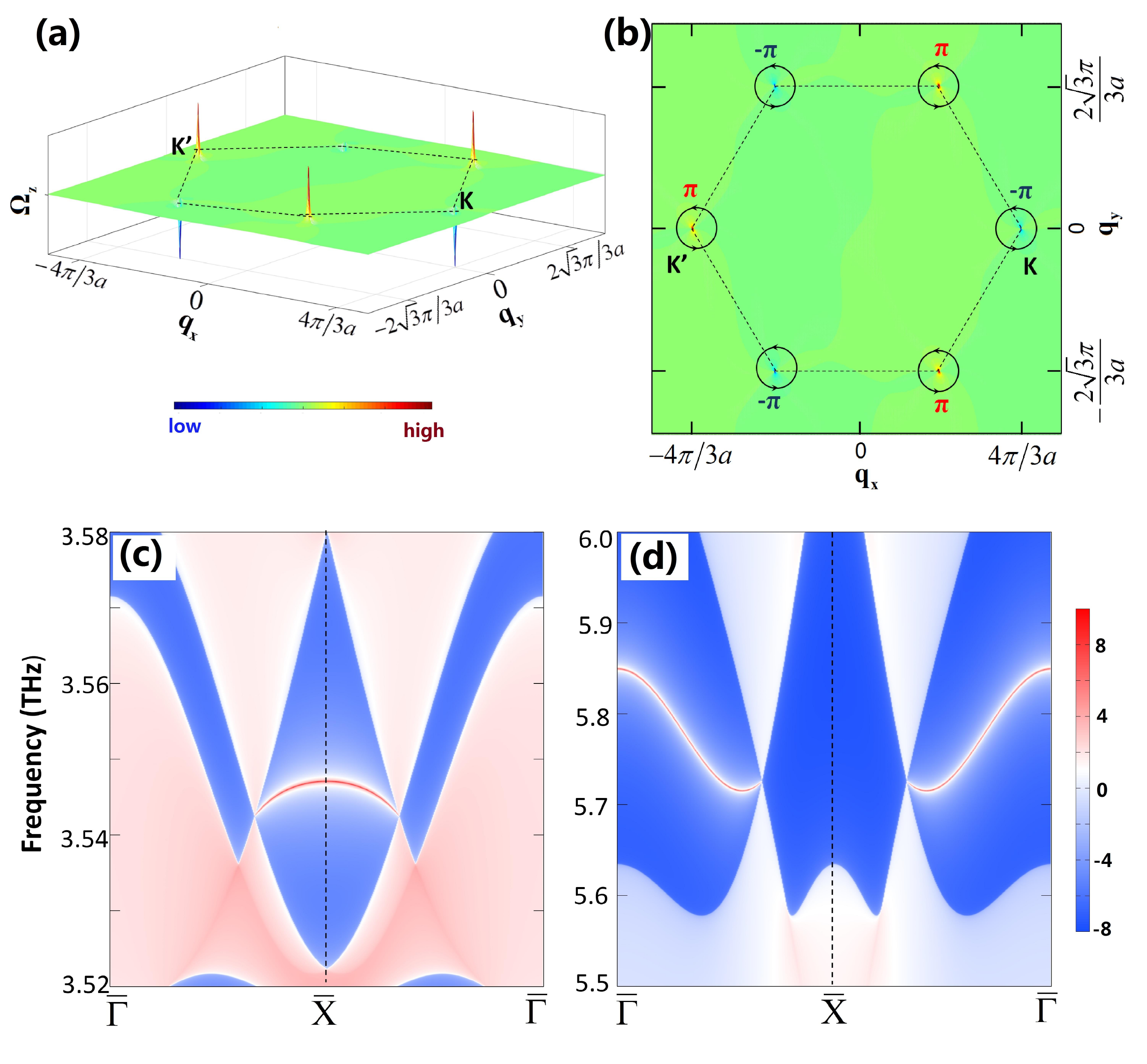}
	\caption{The phonon Berry curvature and local density states (LDOS). The (a) side and (b) top views of Berry curvature distributions of CrI$_3$ with the Dirac frequency $\omega = 3.542$ THz.  The first BZ is marked by the black dashed lines. The quantized phonon Berry phases of $-\pi$ and $\pi$ for two distinct valleys $K$ and $K'$ are also labeled in (b), respectively.  The LDOS projected on edges of semi-infinite nanoribbons of (c) CrI$_3$ and (d) YGaI along the zigzag direction. The nontrivial edge states terminated at the projection of phonon Dirac cones are clearly visible.
\label{edge}}
\end{figure}

It is well known that the Berry curvature is analogical to a magnetic field in momentum space. To reveal the topological features of Dirac phonons, we calculated the phonon Berry curvature using the Wannier TB Hamiltonian constructed from real-space force constants.\cite{TOGO20151} As a representative, we focus on the ideal Dirac phonon of CrI$_3$ with the frequency $\omega = 3.542$ THz.  As shown in \ref{edge}(a) and \ref{edge}(b), we plot the distributions of Berry curvature in momentum space.  As expected, the nonzero ${\Omega}_z(\mathbf{q})$ diverges at $K$ (or $K'$) valleys and rapidly vanishes away from  $K$ (or $K'$) valleys. The distinct valleys $K$ and $K'$ exhibit the opposite values of ${\Omega}_z(\mathbf{q})$. Hence, the integral of Berry curvature ${\Omega}_z(\mathbf{q})$ in whole BZ must be zero, which is in accordance with the coexistence of $\mathcal{IT}$ symmetry.  For a closed path $\ell$ encircling the phonon Dirac point, we have numerically calculated the Berry phase $\gamma_{\ell}$ [see \ref{edge}(b)]. The results show that the nontrivial Berry phase $\gamma_{\ell}$ is precisely quantized to $-\pi$ and $\pi$ around the $K$ and $K'$ valleys, respectively. In electron systems, the nonzero valley Berry curvature will induce an anomalous velocity in the transverse direction when applying an in-plane electric field, leading to a valley electron Hall effect.\cite{PhysRevLett.99.236809} The valley electron Hall effect has been observed in graphene superlattices\cite{Gorbachev448} or monolayer MoS$_2$.\cite{Mak1489} Similarly, the quantized valley phonon Berry phase can also bring in many exotic topological quantum phenomena, such as a valley phonon Hall effect.\cite{PhysRevLett.105.225901} Analogous to electrons, a valley phonon Hall effect should be observed in the presence of an in-plane gradient strain field.\cite{PhysRevLett.96.155901,PhysRevLett.113.265901} The opposite values of Berry curvatures of two valleys induce the opposite transverse phonon currents, which can be used to detect the valley Dirac phonons.

The topological Dirac phonons with quantized Berry phase will lead to the nontrivial edge states. Hence, we calculate the local density of states (LDOS) of phonons, and the iterative Green's function method with the Wannier TB Hamiltonian\cite{Sancho1984} is employed as implemented in the WANNIERTOOLS package.\cite{WU2017} As shown in \ref{edge}(c) and \ref{edge}(d), we illustrate the LDOS projected on zigzag edges of semi-infinite nanoribbons for CrI$_3$ and YGaI, respectively. Here, we focus on the LDOS around the Dirac frequencies $\omega = 3.542$ THz for CrI$_3$ and $\omega = 5.725$ THz for YGaI.  As expected, the topological edge states are terminated at the projections of phonon Dirac cones due to the global topology below the phonon Dirac point. The topologically nontrivial edge states are clearly visible in phonon frequency gaps. The absence of overlap between topological edge states and bulk states leads them to be easily detectable by surface sensitive probes such as electron energy loss spectroscopy and helium scattering.\cite{PhysRevLett.121.035302}

In summary, we have presented a recipe that leads to Dirac phonon states with quantized valley Berry phase in 2D hexagonal lattices. Using this recipe, we show that candidates with the $C_3$ symmetry at corners of the hexagonal BZ host robust valley Dirac phonons, which are guaranteed to remain intact with respect to perturbations. Since phonons are typically spinless, magnetisms of crystals cannot break the $\mathcal{{T}}$ symmetry of phonons. Using first-principles calculations, we propose that Dirac phonon states widely exist in 2D hexagonal materials with different chemical compositions. CrI$_3$ and YGaI are representative materials to show such special topological features of Dirac phonons. Of particular importance is the monolayer CrI$_3$, an important 2D magnetic semiconductor with attractive applications in spintronics, providing an ideal 2D candidate to investigate topological phonons in experiments. The quantized Berry phase of $-\pi$ and $\pi$ respectively at the $K$ and $K'$ valleys are confirmed, and the phonon edge states are visibly terminated at the projections of phonon Dirac cones. Our findings demonstrate that 2D hexagonal lattices with attractive valley Dirac phonons is a powerful recipe for extending valley quantum phenomena for future research.

~~\\
~~\\
\textbf{ASSOCIATED CONTENT}\\
\textbf{Supporting Information}\\
Detailed results of other compounds and a comparison between the ferromagnetic and nonmagnetic phases of the phonon spectra of CrI$_3$, including Figures S1-S4.

~~~\\
~~\\
\textbf{AUTHOR INFORMATION}\\
\textbf{Corresponding Authors}\\
$^*$E-mail: rcwang@cqu.edu.cn (R.W)\\
$^*$E-mail: xuh@sustc.edu.cn (H.X)\\
\\
\textbf{Notes}\\
The authors declare no competing financial interest.

\begin{acknowledgement}

This work is supported by the National Natural Science Foundation of China (NSFC, Grant Nos.11674148 and 11304403), the Guangdong Natural Science Funds for Distinguished Young Scholars (No. 2017B030306008), the Fundamental Research Funds for the Central Universities of China (Nos. 106112017CDJXY300005 and cqu2018CDHB1B01), and Science Challenge Project (No. TZ2016003).

\end{acknowledgement}


%

\newpage
\providecommand*{\mcitethebibliography}{\thebibliography}
\csname @ifundefined\endcsname{endmcitethebibliography}
{\let\endmcitethebibliography\endthebibliography}{}

\newpage

\textbf{For TOC only}\\
~~~\\
~~\\
	\centering
    \includegraphics[width=8.5cm,height=5cm]{TOC.pdf}

\end{document}